%% file: main_revtex.tex
\documentclass[%
 reprint,
 amsmath,amssymb,
 aps,
]{revtex4-2}

\usepackage{graphicx}
\usepackage{dcolumn}
\usepackage{bm}
\usepackage{hyperref}

\renewcommand{\equationautorefname}{Eq.}
\makeatletter
\def\Hy@autoref@equation#1{%
  \begingroup
  \edef\@tempa{\endgroup
    \noexpand\hyperref[#1]{\equationautorefname~(\noexpand\ref*{#1})}}%
  \@tempa
}
\makeatother

\usepackage{siunitx}
\usepackage{physics}
\usepackage[english]{babel}
\renewcommand{\selectlanguage}[1]{}
\input{mathcommands}

\begin{document}

\preprint{APS/123-QED}

\title{Mitigating nonlinear transduction noise\\in high-cooperativity cavity optomechanics}
\author{Daniel Allepuz-Requena}
\email{dalre@dtu.dk}
\author{Zohran Ali}
\author{Dennis Høj}
\author{Yingxuan Chen}
\author{Luiz Couto Correa Pinto Filho}
\altaffiliation{Present address: Danish Fundamental Metrology (DFM), Kogle Alle 5, 2970 Hørsholm, Denmark}
\author{Alexander Huck}
\author{Ulrik L. Andersen}
\email{ulrik.andersen@fysik.dtu.dk}
\affiliation{Center for Macroscopic Quantum States (bigQ), Department of Physics, Technical University of Denmark, Fysikvej, 2800 Kongens Lyngby, Denmark}

\date{\today}

\begin{abstract}

\input{0_abstract}
\end{abstract}

\maketitle


\section{Introduction}
\input{1_introduction.tex}
\section{The nonlinear transduction regime}
\input{2_nonlinear_regime}
\section{Membrane-in-the-middle of a microcavity}
\input{3_MiM}

\section{Observation of third order thermal intermodulation noise}
\input{4_TIN}
\section{Position readout in the nonlinear regime}
\input{5_nonlinear_readout}
\section{Conclusions}
\input{6_conclusions}
\section*{Acknowledgments}
\input{7_acknowledgements}
\bibliography{references, extra}

\end{document}


\preprint{APS/123-QED}

\title{Supplementary Information for ``Mitigating nonlinear transduction noise\\in high-cooperativity cavity optomechanics"}
\author{Daniel Allepuz-Requena}
\email{dalre@dtu.dk}
\author{Zohran Ali}
\author{Dennis Høj}
\author{Yingxuan Chen}
\author{Luiz Couto Correa Pinto Filho}
\altaffiliation{Present address: Danish Fundamental Metrology (DFM), Kogle Alle 5, 2970 Hørsholm, Denmark}
\author{Alexander Huck}
\author{Ulrik L. Andersen}
\email{ulrik.andersen@fysik.dtu.dk}
\affiliation{Center for Macroscopic Quantum States (bigQ), Department of Physics, Technical University of Denmark, Fysikvej, 2800 Kongens Lyngby, Denmark}

\date{\today}

\maketitle
\onecolumngrid
\section{Determination of the single-photon optomechanical coupling in the nonlinear transduction regime}
In this section, we give a detailed description of how the single-photon optomechanical rate $g_0$ is obtained from the visible modulation of the cavity reflection or transmission spectrum by mechanical motion. This is done by comparing the amplitude of the mechanical oscillation derived from the modulated resonance with the expected amplitude distribution of a thermal state. The slow-varying quadratures $X(t)$ and $Y(t)$ of a high-quality ($\Omega_m \gg \Gamma_m$) mechanical resonator follow a Ornstein-Uhlenbeck. The quadratures are found to be zero-centered normally distributed with variance $\ev{X^2}=\ev{Y^2}= {\kB T}/{m\Omega_m^2}$. Thus, the amplitude of the oscillation $A=\sqrt{X^2+Y^2}$ follows a Rayleigh distribution of mean:
\begin{equation}
    \ev{A} = \sqrt{\frac{\pi}{2}} \sqrt{\frac{\kB T}{m \Omega_m^2}} = \xzp\sqrt{\pi} \sqrt{\frac{\kB T}{\hbar \Omega}} \simeq \xzp \sqrt{\pi n_\th}.
    \label{eq:mean_amplitude}
\end{equation}
First, the oscillation amplitude $A$ in meters of the mode of interest by fitting a Lorentzian modified by sinusoidal detuning fluctuations:
\begin{equation}
    T(t) \propto \frac{1}{1+\qty[\nu(t)+\sum_i \alpha_i \cos{\qty(\Omega_i t + \phi_i)} ]^2}.
\end{equation}
Repeated measurement of the optical resonance and fitting we obtain an estimate of:
\begin{equation}
    \ev{\alpha} = \frac{2 G}{\kappa} \ev{A}.
\end{equation}
Together with \autoref{eq:mean_amplitude}, we find the following expression for $g_0$:
\begin{equation}
    g_0 = \frac{\kappa \ev{\alpha}}{2 \sqrt{\pi n_\th}}. 
\end{equation}
In the experiment, we assume the mechanical modes to be in their thermal state by operating the system with on-resonance cooperativites of $C/\nth\approx10^{-3}$, therefore reducing 
the optical damping rate below $1\unit{\hertz}$, while the scanning laser spends below $1\unit{\ms}$ within the resonance's linewidth.
\section{Off-resonant ringdown and nonlinear damping}
Ringdown measurements are done using homodyne detection on the reflection of a probe beam far from the cavity resonance $\Delta/\kappa \approx -10^{5}$. Approximately 1mW of power is sent through the lower-transmissivity mirror (glass substrate 100ppm). The collection efficiency of the light reflected is extremely low due to the small reflectivity of the membrane (1.5\%) and the small transmissivity of the mirror. The homodyne detection scheme is incapable of resolve thermal motion, but capable of detecting large amplitude motion. The resonator is mechanical excited prior to recording the ringdown by tuning the probe laser to the blue-side of the cavity. The resulting amplitude is large enough to push the mechanical resonator past its linear regime. At this point, a string or membrane resonator is better described as a Duffing oscillator with an extraordinary damping proportional to the square of the amplitude~\cite{hoj_development_2021}:
\begin{equation}
    \frac{\mathrm{d}^2 u_m(t)}{\mathrm{d} t^2}+\left(\Gamma_{\text {lin}}+\Gamma_{\text {nlin}} u_m^2\right) \frac{\mathrm{d} u_m}{\mathrm{~d} t}+\left[\omega_m^2+u_m^2(t) \alpha_{\text {eff }}\right] u_m(t)=0,
\end{equation}
where $u_m$ is the mechanical displacement, $\Gamma_{\text{lin}}=\Gamma_m$ is the damping rate in the linear regime, $\Gamma_{\text{nlin}}$ is a nonlinear damping rate and $\alpha_\text{eff}$ is a term depending on the geometry and stress of the resonator. From the previous equation, we derive the following model for the time-evolution of the energy ($E$) of the resonator. 
\begin{equation}
    \dot{E} = -(\Gamma_m + \beta E) E, 
\end{equation}
where $\beta$ characterizes the nonlinear damping. The solution to this differential equation is:
\begin{equation}
    E(t) = \frac{\Gamma_m E_0}{(\Gamma_m + \beta E_0) \exp{\Gamma_m t} - \beta E_0}.
    \label{eq:duffing_ringdown}
\end{equation}
For $\beta=0$, we recover the usual exponential decay $E(t)=E_0 \exp{-\Gamma_m t}$. We use \autoref{eq:duffing_ringdown} (with an additional offset modeling imprecision noise) to fit our ringdown measurements and obtain the quality factor of the resonator.

\section{Power spectral density of direct detection in the presence of frequency noise}
In this section, we introduce a model for a dispersive cavity optomechanical system measured through direct detection of the cavity. This model is used in the main text to fit the measured spectra. The observable measured by direct detection is the output photon rate $\opsoutdagger_\tran\opsout_\tran$ or equivalently the associated amplitude quadrature $\X^\out_\tran=\qty(\opsoutdagger_\tran + \opsout_\tran)/\sqrt{2}$ (arbitrarily setting the intracavity field as the phase reference $\alpha \in \mathbb{R}$). This means that the only relevant Fourier-transformed Langevin equations for the linearized interaction are the intracavity amplitude~\cite{bowenQuantumOptomechanics2016}:
\begin{equation}
    \X(\omega) = \chi_c^X(\omega) \sqrt{n_\cav} \qty(\Delta(\omega)-G\q(\omega)) + \sum_{i=\drive,\tran,\loss} \sqrt{\frac{\kappa_i}{2}} \qty[\chi_c(\omega) \opsin_i(\omega) + \chi_c^*(-\omega) \opsindagger_i(\omega)],
\end{equation}
where:
\begin{equation}
    \chi_c^X(\omega) = \i\qty[\chi_c(\omega) - \chi_c^*(-\omega)]/\sqrt{2},
\end{equation}
is defined from the usual cavity susceptibility:
\begin{equation}
    \chi_\cav \deq 1/\qty(\kappa/2 -\i\Delta -\i\omega)
\end{equation}.
The mechanical fluctuations:
\begin{equation}
    \q(\omega) = \chi_m(\omega) \qty[\F_\th (\omega) - \hbar G \sqrt{n_\cav}\sqrt{2}\X(\omega)].
\end{equation},
where $\chi_m(\omega) = 1/m (\Omega_m^2 - \omega^2 -\i \omega \Gamma_m)$. We solve the linear system for $\X(\omega)$ as a function of noise sources:
\begin{align*}
    \X(\omega) = \frac{1}{1-\hbar G^2 n_\cav \sqrt{2} \chi_c^X(\omega) \chi_m(\omega)} \Bigg[ &\chi_c^X(\omega) \sqrt{n_\cav} \Delta(\omega) \\
    &- G \sqrt{n_\cav}\chi_c^X(\omega) \chi_m(\omega)\F_\th (\omega)\\
    & +  \sum_{i=\drive,\tran,\loss} \sqrt{\frac{\kappa_i}{2}} \qty[\chi_c(\omega) \opsin_i(\omega) + \chi_c^*(-\omega) \opsindagger_i(\omega)] \Bigg].
\end{align*}
The amplitude fluctuations of the transmitted field can be found through the input-output relation:
\begin{equation}
    \X^\out_\tran = \frac{\opsin_\tran+\opsindagger_\tran}{\sqrt{2}} - \sqrt{\kappa_t} \X,
\end{equation}
and obtain the double-sided power spectral density of $\X^\out_\tran$:
\begin{align}
    \PSD{\X^\out_\tran} \qty(\omega) =&\frac{{\kappa_\tran}}{\abssquared{1-\hbar G^2 n_\cav^2 \sqrt{2} \chi_c^X(\omega) \chi_m(\omega)}} \Bigg[  \abssquared{\chi_c^X(\omega)} {n_\cav} \PSD{\Delta}\qty[\omega] 
    + G^2 {n_\cav} \abssquared{\chi_c^X(\omega) \chi_m(\omega)} \PSD{\F}^\th \qty[\omega]
     + \frac{\kappa -\kappa_\tran}{2} \abssquared{\chi_c^*(-\omega)}\Bigg] \\
    &+ \abssquared{\frac{1}{\sqrt{2}} - \frac{\kappa_\tran \chi_c^*(-\omega)  /2}{1-\hbar G^2 n_\cav^2 \sqrt{2} \chi_c^X(\omega) \chi_m(\omega)}}.
    \label{eq:Xout_PSD}
\end{align}
Note that the last term (second line) arises from the contribution $\qty({\opsin_\tran+\opsindagger_\tran})/{\sqrt{2}}$ in the input-output relation. If the transmitted beam is sent directly to a photodetector measuring with total efficiency $\eta_\mathrm{det}$ (that takes into account losses in the collection path and quantum efficiency of the photodiode), the symmetrized photocurrent\footnote{The double-sided symmetrized power spectral density $\sPSD{I} \qty(\omega) \deq \qty(\PSD{I}\qty(\omega) + \PSD{I}\qty(-\omega))/2$ is related to the single-sided power spectral density used in the main text by $\mPSD{I}\qty(\omega) \deq 2 \sPSD{I} \qty(\omega)$} PSD is:
\begin{equation}
    \sPSD{I}\qty(\omega) = \eta_\mathrm{det} \abs{\bar{s}_\tran^\out}^2 \qty[1-\eta_\mathrm{det} + 2 \eta_\mathrm{det}\sPSD{\X^\out_\tran} \qty(\omega) ],
\end{equation}
where $\abs{\bar{s}_\tran^\out}^2$ is the average photon rate transmitted through the cavity. The measurement can be calibrated to shot-noise units by measuring a coherent state with equal photon rate (same average photocurrent signal):
\begin{equation}
        \sPSD{I}^{\mathrm{SNU}}\qty(\omega) = 1-\eta_\mathrm{det} + 2 \eta_\mathrm{det}\sPSD{\X^\out_\tran} \qty(\omega) 
\end{equation}
\subsection{Noise minimum}
Fig.~4(b) in the main text shows a prominent noise reduction at the natural oscillator frequency. This occurs regardless of the spring shift caused by dynamical backaction. We explain this with a simplified model that considers some extraordinary fluctuations in the cavity amplitude quadrature:
\begin{equation}
    X(\omega) = \chi_\cav^X (\omega) G q(\omega) + X_\noise(\omega)
    \label{eq:X_simple}
\end{equation}
where the term $X_\noise$ represents an amplitude noise of general origin, which could be quantum noise from the input channels, classical detuning noise independent of $q$, or nonlinear noise. In this case, the output amplitude fluctuations are:
\begin{equation}
     \PSD{\X^\out_\tran} \qty(\omega) \propto {\abs{1-\hbar G^2 n_\cav^2 \sqrt{2} \chi_c^X(\omega) \chi_m(\omega)}}^{-2} \PSD{X_\noise}(\omega),
\end{equation}
where we have neglected the thermal force. Assuming a frequency independent amplitude noise, a high quality factor $\Gamma_m\simeq0$ and a fast cavity $\chi_c(\omega)\simeq\chi_c(\Omega_m)$, it can be shown that $\PSD{\X^\out_\tran}$ has two relative maxima at the $\pm\Omega_\eff$ predicted by DBA (optical spring effect) and three relative minima, at $\omega=0,\pm\Omega_m$. The relative minima can reach values below unity, corresponding to reduction of amplitude noise, which remarkably occurs at the natural mechanical frequency regardless of the effective oscillation frequency. 
\section{Laser noise reduction with a filter cavity}
\begin{figure}[t]
    \centering
    \includegraphics[]{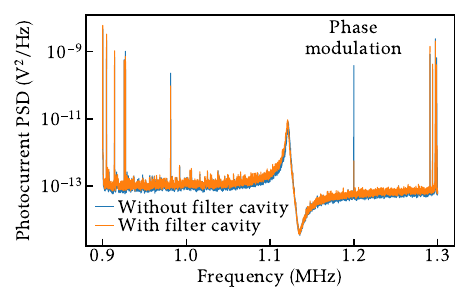}
    \caption{Optomechanical system transmission photocurrent PSDs measured at similar input power and detuning without (blue) and with (orange) a filter cavity. The phase modulation seen at 1.2MHz has been added with an EOM prior to the filter cavity. The reduction of this signal by the filter cavity showcases its ability to remove laser noise.}
    \label{fig:filter_cavity}
\end{figure}
A $\SI{10.9}{\centi\meter}$ linear Fabry-Perot cavity with linewidth $\SI{73}{\kilo\hertz}$ is used to remove laser noise above shot noise. \autoref{fig:filter_cavity} shows how the cavity is capable of reducing a phase modulation added with an electro-optical modulator (EOM) by more than three orders of magnitude. Despite this, the spectrum of the optomechanical system's output remains the same, indicating that the extraneous noise is not dominated by laser noise.

\bibliography{references, extra}

%% file: mathcommands.tex
\DeclareMathOperator{\sign}{sign}

\newcommand{\seesupp}{\cite{suppl}}

\newcommand{\mPSD}[1]{S_{{#1}}}

\renewcommand{\i}{\mathrm{i}}

\renewcommand{\th}{\mathrm{th}}

\newcommand{\cav}{\mathrm{c}}

\newcommand{\inn}{\mathrm{in}}

\newcommand{\eff}{\mathrm{eff}}

\newcommand{\deq}{:=}

\newcommand{\laser}{\mathrm{l}}

\newcommand{\xzp}{x_\mathrm{zp}}
\newcommand{\nth}{n_\th}

\newcommand{\dnu}{\delta\nu}

\newcommand{\calS}{\mathcal{S}}

%% file: 0_abstract.tex
Coupling mechanical motion to an optical resonator enables displacement measurements approaching the standard quantum limit (SQL). However, increasing the optomechanical coupling strength will inevitably lead to probing of the nonlinear response of the optical resonator. Thermal intermodulation noise (TIN) arising from the nonlinear mixing of thermomechanical motion can further increase the imprecision well above the SQL and has hitherto been canceled up to second order of nonlinearity via operation at the ``magic detuning". In this work, we record the output of a membrane-in-the-middle microcavity system operating at room temperature and achieving high cooperativity, $C>n_\th$, and apply a nonlinear transform that removes all orders of TIN, improving the mechanical signal-to-noise ratio by nearly 10~dB. Our results can be applied to experiments affected by third-order TIN, which we expect to be the dominating intrinsic source of noise in high-cooperativity room-temperature cavity optomechanical systems.

%% file: 1_introduction.tex
Cavity optomechanics has enabled a series of remarkable demonstrations of quantum behavior in massive mechanical systems, including ponderomotive squeezing~\cite{safavi-naeini_squeezed_2013, purdy_strong_2013} and ground-state preparation~\cite{chan_laser_2011,rossiMeasurementbasedQuantumControl2018}. Recently, significant effort has been directed toward extending these achievements to room temperature, where the practicality and scalability of optomechanical platforms can be fully exploited~\cite{barzanjehOptomechanicsQuantumTechnologies2022}. In particular, suspended membrane resonators with engineered phononic structures~\cite{tsaturyan_ultracoherent_2017,engelsen_ultrahigh-quality-factor_2024,hoj_ultracoherent_2024} have emerged as a promising route, exhibiting quantum behavior at room temperature, including ponderomotive squeezing~\cite{huang_room-temperature_2024-1} and motional sideband asymmetry~\cite{xia_motional_2024}. Yet, unconditional ground-state preparation at room temperature remains elusive.\par

A main obstacle arises from the fact that accessing the quantum regime requires large optomechanical cooperativity, which in turn demands either high intra-cavity photon number or large single-photon coupling. The latter can drive the system into the nonlinear transduction regime, where thermomechanical motion modulates the cavity detuning by an amount comparable to the cavity optical linewidth. Suspended membrane resonators are more susceptible to this regime due to their dense mechanical spectrum compared to levitated systems. In this regime, the standard linearized treatment of cavity optomechanics fails: the nonlinear cavity response causes mixing between thermomechanical modes, creating thermal intermodulation noise (TIN) and its accompanying backaction (TINBA)~\cite{leijssen_nonlinear_2017, fedorov_thermal_2020,pluchar_thermal_2023}. Even when second-order TIN is eliminated by operating at the ``magic detuning", higher-order mixing persists and can dominate the displacement imprecision in high-cooperativity room-temperature experiments. \par
Up to now, TIN has been canceled only to second-order~\cite{huang_room-temperature_2024-1, xia_motional_2025}, and no experimental platform has operated deeply enough into the nonlinear regime to probe or mitigate higher-order TIN. This gap has become increasingly important as state-of-the-art membrane-in-the-middle systems---particularly those incorporating ultracoherent density-phononic membranes---achieve cooperativities approaching or even exceeding the thermal occupation $C\geq\nth$. In this emerging regime, third-order TIN is expected to be the leading intrinsic nonlinearity, yet it has never been directly identified or mitigated.\par
In this work, we close this gap. We present a membrane-in-the-middle microcavity that operates deep in the nonlinear transduction regime at room temperature, enabling us to directly observe and characterize third-order TIN. Using a combination of spectral analysis and correlation measurements, we trace narrow features in the transmission spectrum to specific triple-mode mixing processes among thermomechanical modes below the phononic bandgap. This constitutes, to our knowledge, the first experimental identification of third-order TIN in any cavity optomechanical system. \par

Beyond identifying the noise, we introduce and implement a nonlinear position-reconstruction protocol that inverts the full cavity response and therefore removes all orders of TIN from the measurement record. This reconstruction restores linear position readout even under large detuning excursions, eliminating the dominant technical limitation of high-cooperativity room-temperature optomechanics. By applying this method to our microcavity system, we demonstrate an almost 10~dB improvement in signal-to-noise ratio and recover a clean mechanical spectrum otherwise obscured by nonlinear mixing. This capability opens a pathway toward accessing the quantum backaction-dominated regime at room temperature, where higher-order classical forces—rather than thermal noise—set the ultimate limits on performance.

%% file: 2_nonlinear_regime.tex
\begin{figure}
    \centering
    \includegraphics[width=\linewidth]{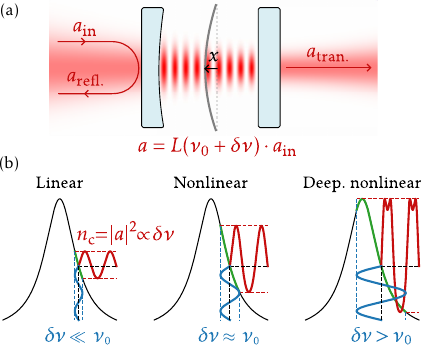}
    \caption{Dispersive transduction of mechanical motion in a membrane-in-the-middle cavity. (a) In the fast-cavity limit, the intracavity field $a$ adiabatically follows relative detuning fluctuations $\delta\nu$ around the equilibrium $\nu_0$. $\delta\nu$ is caused by linear optomechanical coupling of the membrane's displacement $x$ according to Eq.~\eqref{eq:dispersive_coupling}. The reflected $a_\text{refl.}$ or transmitted $a_\text{tran.}$ field are used to infer $a$ and therefore $\delta \nu$. (b) Depending on the magnitude of the detuning fluctuations, their transduction into cavity occupation $n_\cav=\abs{a}^2$ can be approximately linear, nonlinear but unambiguous or, deeply nonlinear and ambiguous.}
    \label{fig:transduction}
\end{figure}
\begin{figure*}[t]
    \centering
    \includegraphics[width=\linewidth]{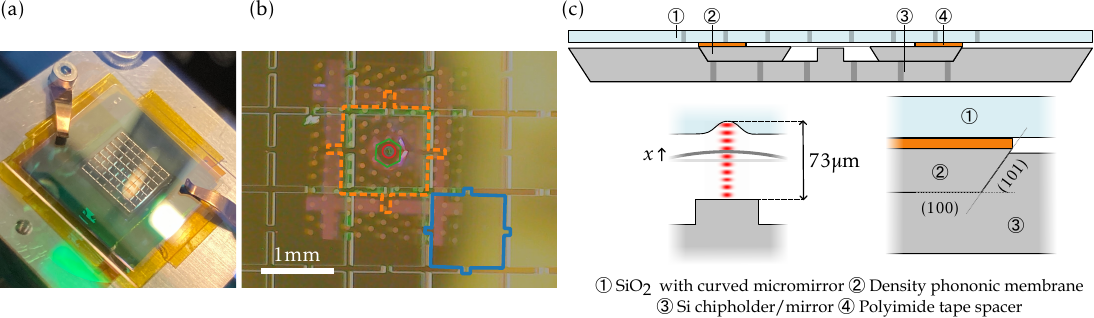}
    \caption{(a) Photograph of the microcavity membrane-in-the-middle  (MIM) system. (b) Vertical optical microscope picture, orange dashed line: bottom mirror phononic cell, blue solid line: top mirror phononic cell, red circle: concave micromirror in top substrate, green hexagon: defect in the density phononic membrane. (c) Schematic profile view of the stack forming the MIM system. The membrane's substrate (label 2) and the bottom mirror (label 3) are etched along the same crystalline planes (100) and (101).}

    \label{fig:MIM}
\end{figure*}
Fig.~\ref{fig:transduction} illustrates how the nonlinear transduction regime arises when cavity frequency fluctuations become comparable to the optical linewidth. In this regime, mechanical motion still perturbs the detuning linearly, but the detuning is mapped nonlinearly onto the intracavity field amplitude and phase. As a result, any attempt to infer position from a linearized measurement of the transmitted or reflected field inevitably mixes higher-order terms of the cavity response, producing thermal intermodulation noise. However, as recently shown in Ref.~\cite{clarke_cavity_2023}, nonlinear transduction does not preclude accurate position measurement: if the full cavity response is taken into account, the true detuning can be reconstructed exactly, and TIN disappears from the measurement record. In this section, we summarize how such reconstruction can be achieved in a direct-detection scheme. 
We start by considering linear dispersive coupling between the cavity and mechanical displacement $x$:
\begin{equation}
    \nu(t) = \nu_0 - \frac{2 G}{\kappa} x(t),
    \label{eq:dispersive_coupling}
\end{equation}
where $\nu=2\Delta/\kappa$ is the detuning $\Delta = \omega_\laser - \omega_\cav$ between the laser and cavity frequency relative to the cavity linewidth $\kappa$. The cavity pull factor $G$ characterizes the strength of the optomechanical coupling. $\nu_0$ is the detuning in the absence of mechanical motion. In the fast-cavity limit, the intracavity field $a$ is taken to adiabatically follow detuning changes with a normalized susceptibility~\cite{fedorov_thermal_2020}:
\begin{equation}
    a = L(\nu) a_\inn = \frac{1}{1-\i \nu} a_\inn,
\end{equation}
where $a_\inn$ is the input field. This leads to the intracavity photon number
\begin{equation}
    n_\cav =\abs{a}^2 = n_{\cav,0} \abs{L(\nu)}^2 = \frac{n_{\cav,0}}{1+\nu^2}
    \label{eq:L_ncav}
\end{equation}
and phase
\begin{equation}
    \varphi = \varphi_0 + \operatorname{Arg}\qty{L(\nu)} = \varphi_0 + \arctan{\nu},
    \label{eq:L_phi}
\end{equation}
where $n_{\cav,0}$ and $\varphi_0$ are the on-resonance intracavity intensity and phase.\par
Our aim is to recover $\nu(t)$, and thereby the mechanical motion via Eq.~\eqref{eq:dispersive_coupling}, using only measurable quantities inferred from the cavity output field. Although neither $n_\cav$ nor $\varphi$ are directly accessible independently, both can be inferred from the field reflected or transmitted from the cavity.\par
Expanding Eq.~\eqref{eq:L_ncav} around $\nu_0$ yields the common linear approximation valid for small detuning fluctuations $\delta\nu(t)=\nu(t)-\nu_0$,
\begin{equation}
    \dnu(t) \simeq \frac{1}{\partial_\nu \abs{L(\nu_0)}^2}\qty [\frac{n_\cav(t)}{n_{\cav,0}} - \abs{L(\nu_0)}^2], 
    \label{eq:linear_readout}
\end{equation}
with a similar result obtained from expanding Eq.~\eqref{eq:L_phi}. The higher order terms in the expansion, $\dnu^n$, give rise to TIN in any linearized  displacement measurement. The dynamical effects caused by the radiation pressure exerted by TIN (TINBA) are always present. However, the additional position readout noise only appears if the linear approximation is made. Instead of expanding and approximating Eq.~\eqref{eq:L_ncav} and Eq.~\eqref{eq:L_phi}, we may directly invert them:
\begin{equation}
    \nu(t) = \sign{\varphi}\abs{\sqrt{\frac{n_{\cav,0}}{n_\cav (t)}-1}}
    \label{eq:nu_t}
\end{equation}
\begin{equation}
    \nu(t) = \tan\qty[\varphi(t)-\varphi_0]
\end{equation}
which take into account the full nonlinear response and are thus immune to all orders of TIN. In a direct-detection scheme where the photocurrent is proportional to $n_\cav(t)$, the symmetry $\abs{L(\nu)}^2=\abs{L(-\nu)}^2$ creates a sign ambiguity. This can be resolved by knowledge of the sign of the phase, which is anti-symmetric on detuning. The requirement of simultaneous amplitude and phase measurements makes this scheme equivalent to the general-dyne protocol described in Ref. \cite{clarke_cavity_2023}. 
Crucially, in many situations the phase measurement is unnecessary. If the instantaneous detuning does not cross zero during the measurement record, the sign of 
$\nu(t)$ remains fixed by the sign of the operating detuning $\nu_0$, and we may use
\begin{equation}
    \nu(t) = \sign{\nu_0}\abs{\sqrt{\frac{n_{\cav,0}}{n_\cav (t)}-1}} \quad \qty(\abs{\dnu(t)}<\abs{\nu_0}\ \forall t),
    \label{eq:non_linear_readout}
\end{equation}
thus foregoing the need of general-dyne detection while retaining the immunity to TIN.\par
In the remainder of this paper, we apply this reconstruction technique to a membrane-in-the-middle microcavity operating deep in the nonlinear transduction regime, enabling us to identify higher-order TIN and demonstrate TIN-free position readout.

%% file: 3_MiM.tex
\begin{figure*}[t]
    \centering
    \includegraphics[width=\textwidth]{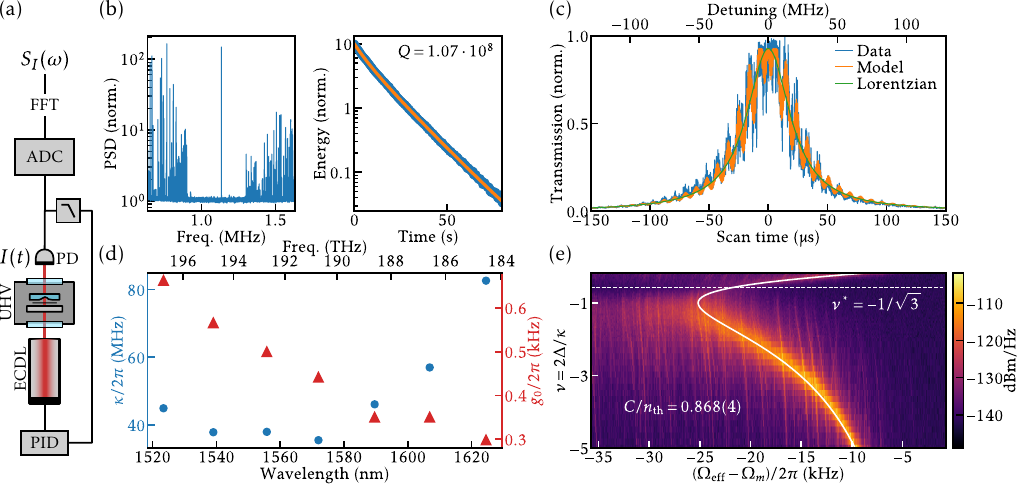}
    \caption{(a) Experimental setup. ECDL: external cavity diode laser, UHV: ultra high vacuum chamber, PD: photodetector, PID: proportional-integral-derivative controller that adjusts the laser's cavity, ADC: analog-to-digital converter used to record time traces. (b) Left: photocurrent power spectral density (PSD). Right: In situ ringdown measurement of the quality factor of the mechanical resonator. Measured energy (blue) is fit with a model with linear and nonlinear damping (orange). (c) Scan of the laser frequency across the optical resonance at $\SI{1572}{\nm}$. The model consists of a Lorentzian resonance with central frequency modulated by two sinusoidal signals of $\SI{1.13}{\mega\hertz}$ (mode of interest) and $\SI{110}{\kilo\hertz}$ (out-of-bandgap mode). (d) Cavity linewidth (blue, left-hand axis) and single-photon optomechanical rate (orange, right-hand axis) at the accessible HG$_{00}$ optical resonances. (e) Photocurrent PSD measured at increasing detuning with constant intracavity power (\SI{1571}{\nano\meter} optical resonance). The change in oscillation frequency is fit with a model of optical spring effect. }
    \label{fig:cavity_char}
\end{figure*}
\begin{figure*}[t]
    \centering
    \includegraphics[]{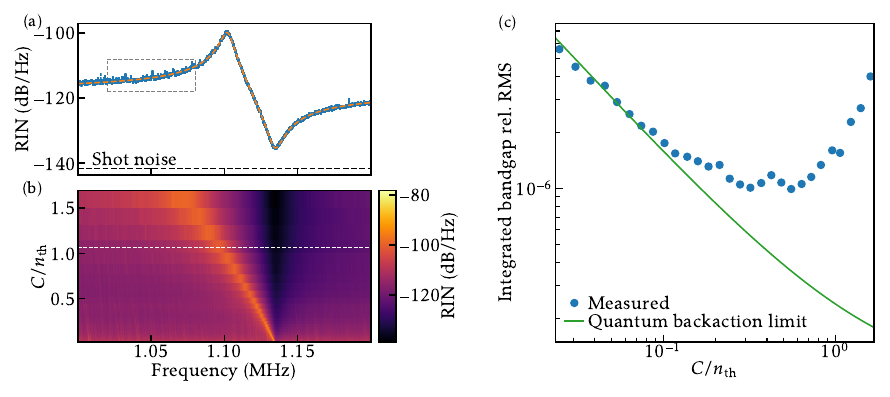}
    \caption{(a) Relative intensity noise (RIN) spectrum of the detected transmission  with $C\simeq\nth$. Orange dashed line: fit with extraneous amplitude noise model. (b) Transmission RIN spectra at the magic detuning at different cooperativities. Dashed white line: spectrum displayed in (a). (c) Relative root mean square (RMS) noise in the frequency range of the membrane mechanical bandgap. Green solid line represents the expected behavior in the absence of technical noise.}
    \label{fig:PSDs}
\end{figure*}
Our Fabry-Perot microcavity optomechanical system, shown in Fig.~\ref{fig:MIM}, has been designed to minimize the cavity length while integrating in the middle a chip carrying a mechanical resonator. The density phononic membrane (DPM) used in our experiment is a $\SI{20}{\nm}$ thick silicon-nitride membrane suspended on a $\SI{1}{\cm^2} \times \SI{500}{\um}$ silicon frame~\cite{hoj_ultracoherent_2024}. In the membrane, a defect in a pillar-defined phononic crystal hosts a high quality factor mechanical mode (from now on referred to as the high-Q mode). In order to achieve cavity lengths below the DPM chip thickness, a protruding ``mesa" mirror has been sculpted into a silicon chip holder (label 3 in Fig.~\ref{fig:MIM}) etched with the negative of the DPM chip (label 2 in Fig.~\ref{fig:MIM}). The top mirror (label 1 Fig.~\ref{fig:MIM}) is made on a glass substrate where a concave feature with a $\approx \SI{270}{\um}$ radius of curvature has been added through laser ablation~\cite{petrak_feedback-controlled_2011, allepuz_requena_strong_2024}. Both mirrors are coated with dielectric distributed Bragg reflectors (SiO$_2$/Ta$_2$O$_5$) with target transmissions of 10ppm (bottom) and 100ppm (top). Contact between the membrane and the top mirror is prevented by a single layer of $\SI{30}{\um}$-thick polyimide tape (label 4 in Fig.~\ref{fig:MIM}), setting a cavity length of $\SI{73}{\um}$.\par
The short cavity length increases optomechanical transduction while minimizing the effects of laser frequency noise. In order to suppress cavity frequency noise, the mirror substrates have been etched with a phononic crystal pattern (PnC)~\cite{saarinen_laser_2023,huang_room-temperature_2024-1}. The PnC creates a vibrational bandgap around the frequency of the high-Q mode. The width of the bandgap has been optimized by finite-element simulations of PnC. The resulting optimal pattern of crosses defining the PnC~\cite{yu_phononic_2014} can be appreciated in Fig.~\ref{fig:MIM}(a) and Fig.~\ref{fig:MIM}(b).\par
We now proceed with characterization of the mechanical resonator, the optical cavity and the strength of the optomechanical coupling. The system is operated at room temperature and pressures of $\approx \SI{5e-8}{}\text{mbar}$. Characterization is done via direct detection of the cavity's transmission, as shown in Fig.~\ref{fig:cavity_char}(a). The photocurrent $I(t)$ is recorded using an analog-to-digital converter. The single-sided photocurrent power spectral density $\mPSD{I}(\omega)$ (from now on photocurrent PSD) is obtained through the fast Fourier transform and Welch's method. The driving laser is an external cavity diode laser (TOPTICA CTL1550) with tunability from $\SI{1510}{\nm}$ to $\SI{1630}{\nm}$, covering approximately 6 times the $\SI{2.05} {\tera\hertz}$ free-spectral range of the microcavity.\par   
The high-Q mechanical mode is identified by operating the system at a large detuning, where the photocurrent fluctuations are approximately proportional to the membrane's thermomechanical motion. The photocurrent PSD presented in the right-hand side of Fig.~\ref{fig:cavity_char}(b) shows the $\Omega_m = \SI{1.13}{\mega\hertz}$ high-Q mechanical mode within the roughly $\SI{300}{\kilo\hertz}$-wide bandgap. Its natural damping rate is measured by a ringdown measurement. The mode is first excited by tuning the laser to the blue-side of the cavity in order to induce self-oscillations and then moved completely off-resonance in order to remove the effects of dynamical backaction. The motion was then measured through homodyne detection on the field reflected off the optical cavity (not shown in Fig.~\ref{fig:cavity_char}(a)). The decay of energy in the mechanical mode is tracked during the span of a minute (right-hand side of Fig.~\ref{fig:cavity_char}(b)) and fitted with a model that includes nonlinear damping at large amplitudes~\seesupp. The observed quality factor of $Q= \Omega_m / \Gamma_m =\SI{1.071(5)e8}{}$ places our resonator deep in the ultracoherent regime, where at room temperature $\Omega_m \simeq 19 \Gamma_m \nth $ and $\Gamma_m \nth = \Gamma_{\th}$ being the thermal mechanical dissipation rate.\par 

The optical cavity transmission spectrum obtained with a fast scan ($\SI{1}{\mega\hertz/\micro\second}$) of the laser frequency across the optical resonance is shown in Fig.~\ref{fig:cavity_char}(c) and reveals operation of our system deep in the nonlinear transduction regime. The Lorentzian optical resonance is distorted in the manner expected from the nonlinear model presented in Eq.~\eqref{eq:L_ncav}. The cavity linewidth is extracted from a fit using this model and the extracted values are presented in Fig.~\ref{fig:cavity_char}(d) for the range of accessible optical cavity modes. Furthermore, the fit also reveals the amplitude of the mechanical modes causing the distortion of the transmission spectrum. The average amplitude of the contributing mechanical modes (in units of relative detuning) can be estimated from repeated scans of the resonance. Together with the linewidth and considering thermal occupation, the magnitude of the single-photon optomechanical coupling rate $g_0=G \xzp$ can be estimated for each mechanical mode~\seesupp. We present the $g_0$ estimations for the high-Q mode in Fig.~\ref{fig:cavity_char}(d) for the accessible optical cavity resonances.\par

The ultra-low mechanical damping rate ($\Gamma_m$) together with the large optomechanical coupling ($g_0$) and low cavity dissipation ($\kappa$) allows our system to reach cooperativities that fulfill $C=4g_0^2 \bar{n}_\cav/\kappa\Gamma_m>n_\th$ where the strength of the quantum backaction force $\mPSD{\text{QBA}}(\Omega_m) \simeq 8 \hbar^2 g_0^2 \bar{n}_\cav / \xzp^ 2 \kappa$ exceeds that of the thermal force $\mPSD{\th}(\Omega_m) \simeq 2 \hbar^2 \Gamma_m n_\th/\xzp^2$. We characterize $C$ by analysis of the optical spring effect. To do this, we realize sweeps of detuning $\nu$ at constant intracavity power $\bar{n}_c$ by locking the laser power to a constant transmission signal. In this manner, any thermal frequency shift caused by optical heating of the membrane remains constant throughout the measurement and we avoid overestimation of the cooperativity (an effect also seen in Ref.~\cite{xia_motional_2024}). The effective mechanical oscillation frequency in the fast-cavity regime ($\kappa\gg\Omega_m$) and at constant average $\bar{n}_\cav$ is given by~\cite{aspelmeyerCavityOptomechanics2014}:
\begin{equation}
    \Omega_\eff = \Omega_m + \Gamma_m C \frac{\nu}{1+\nu^2}.
\end{equation}
Fig.~\ref{fig:cavity_char}(e) shows the result of fitting the expression to a series of PSDs obtained from the photocurrent as a function of detuning $\nu$. The detuning corresponding to each input power is not calibrated prior to fitting. Instead, the initial detuning of the measurement series is left as a free-parameter, while the remaining detunings can be extrapolated from the ratio of the input power to the first input power value of the series. We find that the maximum reachable $C$ depends on the chosen optical cavity mode. Ideally, this limit is given by the intrinsic single-photon cooperativity at each mode $C_0 = C/\bar{n}_\cav$ and the accessible intracavity photon number at the particular wavelength. In practice, operation at large $C$ however is prevented by out-of-bandgap mechanical modes with a large amplitude making the laser-lock to the cavity unstable.\par

We further explore the regime of high cooperativity $C>\nth$ by probing the system at the magic detuning $\nu^* = -1/\sqrt{3}$, where second-order TIN is canceled ($\partial^2_\nu \abs{L(\nu^*)}^2=0$). We chose the optical resonance close to \SI{1571}{\nano\meter}, where the system remains stable up to $C\approx 2\nth$. Operating at $C\simeq\nth$, the photocurrent reveals background noise about 20~dB above shot noise in the vicinity of the high-Q mechanical mode, as shown by the relative intensity noise (RIN) spectrum in Fig.~\ref{fig:PSDs}(a). Measurements taken at further increasing cooperativity reveal how the extraneous background noise increases while the mechanical mode spectrum becomes increasingly non-Lorentzian (Fig.~\ref{fig:PSDs}(b)). The observed behavior of background noise is consistent with the presence of extraneous intracavity amplitude fluctuations generated by cavity frequency noise and included in the model fit in Fig.~\ref{fig:PSDs}(a)~\seesupp. In our experiment, this noise is not removed by cleaning the probe laser using a filter cavity with a linewidth of $\SI{73}{\kilo\hertz}$ before injection into the microcavity, indicating that the laser is not the dominating source of the observed background noise~\cite{suppl} and that the source must be intrinsic to the cavity optomechanical system. At this stage of our experiment, we speculate the sources of cavity frequency noise can be residual thermomechanical motion of the optical cavity mirrors or noise originating in the high-reflectivity  coating~\cite{harry_thermal_2002,chalermsongsak_broadband_2014}.\par
Without extraneous noise, sideband cooling is expected to asymptotically reduce the phonon occupation towards $\bar{n}=\kappa/4\Omega_m\simeq 8$ for our system~\cite{marquardt_quantum_2007}. Instead, in the presence of classical background noise, sideband cooling is frustrated by the quadratic dependence of classical fluctuations on intracavity power as compared to the linear scaling of quantum noise~\cite{aspelmeyerCavityOptomechanics2014}. The RIN integrated over the membrane bandgap is plotted Fig.~\ref{fig:PSDs}(c) as a function of cooperativity. At smaller cooperativities $C / \nth < 0.1$, the recorded noise power follows the behavior expected from the presence of cooling the mechanical motion due to thermal and quantum backaction forces. In contrast to this, in the large cooperativity region $C / \nth > 0.1$, the fluctuations grow in accordance with the presence of an additional classically stochastic radiation pressure.\par
Beyond the broadband noise that prevents our system from reaching the quantum regime, smaller noise peaks can be observed in Fig.~\ref{fig:PSDs}(a) in the region highlighted by the dashed box. In the next section, we will investigate the origin of these peaks.\par

%% file: 4_TIN.tex
\begin{figure}
    \centering
    \includegraphics[]{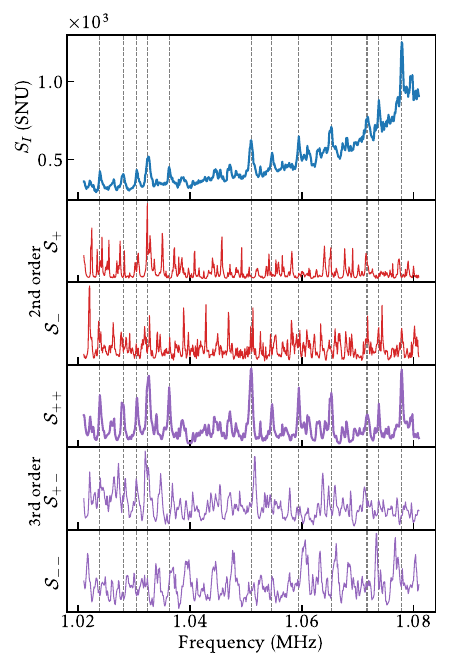}
    \caption{Nonlinearity measures computed from a direct detection spectrum (top plot, dashed region in Fig.~\ref{fig:PSDs}(b)). The nonlinearity measures of second-order (second to third plot) and third-order (fourth to bottom plot) have been normalized to their maximum value in the frequency span displayed.} 
    \label{fig:TIN}
\end{figure}
The spectrum indicated by the dashed box in Fig.~\ref{fig:PSDs}(a) is shown with high resolution in the top of Fig.\ref{fig:TIN}, featuring several sharp noise peaks.
In the following, we first identify the origin of these peaks by comparing the measured RIN spectrum with the spectra of quadratic $\mPSD{\nu^2}$ and cubic $\mPSD{\nu^3}$ detuning fluctuations. Estimating such spectra requires a linear measurement of $\dnu(t)$, which is generally unavailable. One could use the nonlinear $\nu(t)$ reconstruction introduced in \autoref{eq:nu_t}, however, we cannot ensure that the instantaneous detuning remains on the red side when operating the experiment at the magic detuning. Instead, we assume a linear relation between the photocurrent $I(t)$ and $\dnu(t)$ so that $\mPSD{\nu} \propto \mPSD{I}$ (at the magic detuning, this approximation is valid up to third-order of $\dnu$). In order to estimate the spectrum of TIN from $\mPSD{I}$, we define the following quantities as proxies for the power of second-order TIN at a given frequency $\Omega$:
\begin{equation}
    \mathcal{S}_+ (\Omega) \deq \int_{0}^{\Omega} \dd{\omega}\mPSD{I}(\omega)\mPSD{I}(\omega-\Omega)
\end{equation}
\begin{equation}
    \mathcal{S}_- (\Omega) \deq \int_{0}^{\infty} \dd{\omega}\mPSD{I}(\omega)\mPSD{I}(\omega+\Omega).
\end{equation}
These expressions are derived from dividing the relation $\mPSD{\nu^2}(\Omega) \propto \int_{-\infty}^\infty \dd{\omega} \mPSD{\nu}(\omega)\mPSD{\nu}(\Omega-\omega)$~\cite{pluchar_thermal_2023} into contributions of addition ($\mathcal{S}_+$) and subtraction ($\mathcal{S}_-$) of positive frequencies. Intuitively, $\calS_+(\Omega)$ collects the joint power of $\mPSD{I}$ at two frequencies $\Omega_1$ and $\Omega_2$ that fulfill $\Omega=\Omega_1+\Omega_2$, while for $\calS_-(\Omega)$ we have $\Omega=\Omega_1-\Omega_2$. Using the same approach, we can define proxies for the strength of third-order TIN:
\begin{equation}
    \calS_{++}(\Omega) = \int_0^\Omega \!\!\!\!\dd{\omega_1}\!\!\int_0^{\Omega-\omega_1} \!\!\!\!\!\!\!\!\!\!\dd{\omega_2} \mPSD{I}(\omega_1) \mPSD{I}(\omega_2) \mPSD{I}(\Omega-\omega_1-\omega_2),
\end{equation}
\begin{equation}
    \calS_{+-}(\Omega) = \int_0^\infty\!\!\!\!\dd{\omega_1}\!\!\int_0^{\Omega+\omega_1} \!\!\!\!\!\!\!\!\!\!\dd{\omega_2} \mPSD{I}(\omega_1) \mPSD{I}(\omega_2) \mPSD{I}(\Omega+\omega_1-\omega_2),
\end{equation}
and
\begin{equation}
    \calS_{--}(\Omega) = \int_0^\infty\!\!\!\!\dd{\omega_1}\!\!\int_0^{\infty} \!\!\!\!\!\!\dd{\omega_2} \mPSD{I}(\omega_1) \mPSD{I}(\omega_2) \mPSD{I}(\Omega+\omega_1+\omega_2), 
\end{equation}
corresponding to the unique combinations of $\Omega=\Omega_1\pm\Omega_2\pm\Omega_3$ generated in third-order mixing.
Using the measured $S_I(\omega)$, we compute these quantities and present them in Fig.~\ref{fig:TIN} together with the measured spectrum in the displayed spectral region (top). Qualitatively, we observe a correlation in the distribution of peaks between the measured spectrum and the third-order nonlinearity measure of triple frequency addition $\calS_{++}$. In particular, the shared peak at approximately $\SI{1.051}{\mega\hertz}$ corresponds to three prominent peaks below the mechanical bandgap with frequencies $\SI{103}{\kilo\hertz}$, $\SI{296}{\kilo\hertz}$ and $\SI{652}{\kilo\hertz}$.\par
Although the similarity between the estimated and measured spectrum is a good indicator of the presence of third-order TIN, it does not take into account the correlation that must exist between the original linear components and the resulting TIN component. Such correlations can be found by decomposing the photocurrent into individual frequency components:
\begin{equation}
    I(t) = \sum_k  I_k(t) \simeq \sum_k X_k(t) \cos\Omega_k t + Y_k(t) \sin\Omega_k t, 
\end{equation}
where $X_k(t)$ and $Y_k(t)$ are slowly varying quadratures compared to $\Omega_k$. If the component $k=4$ originates in third-order mixing of $k=1,2,3$ so that $\Omega_4 = \Omega_1 + \Omega_2 + \Omega_3$, then, according to the Taylor expansion of Eq.~\eqref{eq:L_ncav}, its quadratures fulfill:
\begin{equation}
    X_4 = \beta_X \cdot \qty( X_1 X_2 X_3 - X_1 Y_2 Y_3 - X_2 Y_1 Y_3 - X_3 Y_1 Y_2 )
    \label{eq:X++}
\end{equation}
\begin{equation}
    Y_4 = \beta_Y \cdot \qty(  X_1 X_2 Y_3 + X_1 X_3 Y_2  + X_2 X_3 Y_1 -Y_1 Y_2 Y_3),
    \label{eq:Y++}
\end{equation}
The proportionality factors $\beta_{X/Y}$ (in principle identical) depend on the relative detuning and the responsivity of the photodetector.\par
\begin{figure}
    \centering
    \includegraphics[]{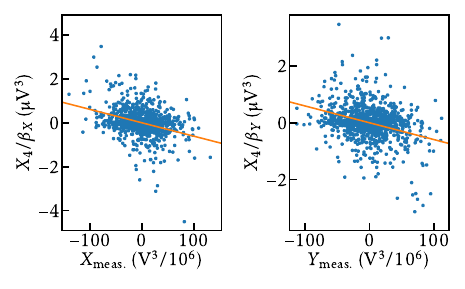}
    \caption{Correlation between the recorded quadratures $X_\text{meas.}$ (left) and $Y_\text{meas.}$ (right) of a third-order TIN peak and their values expected from the measured quadratures of the thermomechanical peaks that originate it (calculated using Eq.~\eqref{eq:X++} and \eqref{eq:Y++}. The solid orange line represents the total least squares fit used to estimate $\beta_{X/Y}$.} 
    \label{fig:TIN_correlation}
\end{figure}
To confirm that the peak at $\SI{1.051}{\mega\hertz}$ in $S_I(\omega)$ arises from mixing of these particular frequency components, we record the photocurrent for $\SI{1}{\second}$ and demodulate both the three thermomechanical modes and the candidate TIN peak to obtain the slowly varying quadratures $X_\text{meas.}$ and $Y_\text{meas.}$. Fig.~\ref{fig:TIN_correlation} shows scatter plots of the measured TIN quadratures versus the values predicted using Eq.~\eqref{eq:X++} and Eq.~\eqref{eq:Y++}. The Pearson correlation, the covariance relative to the product of standard deviations, of -0.30 ($X$) and -0.26 ($Y$) together with the consistent proportionality factors $\beta_X=\SI{-164\pm14}{\volt^{-2}}$ and $\beta_Y=\SI{-168\pm15}{\volt^{-2}}$ support our hypothesis that the noise peaks originate from third-order mixing. At the magic detuning, we have $\partial_\nu \abs{L(\nu^*)}^2 > 0$ and $\partial_\nu^3 \abs{L(\nu^*)}^2 < 0$, explaining the negative proportionality factors $\beta_X$ and $\beta_Y$ observed in our measurement.    

%% file: 5_nonlinear_readout.tex
\begin{figure}[t]
    \centering
    \includegraphics[]{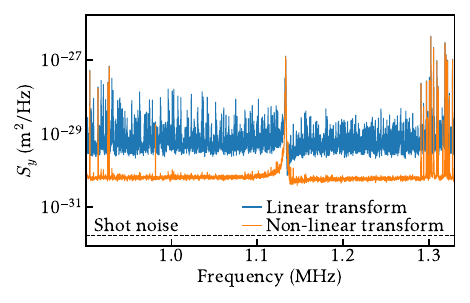}
    \caption{Estimate of the mechanical noise power density assuming linear optomechanical transduction (blue) and considering nonlinear transduction (orange). The remaining peaks below \SI{1.1}{MHz} and above \SI{1.3}{MHz} stem from mechanical modes in the optical cavity.}
    \label{fig:non_linear_readout}
\end{figure}

Third-order TIN is present in any linearized measurement at the magic detuning but vanishes at $\nu=\pm 1$. As suggested previously, only the consideration of the nonlinear cavity response can yield a TIN-free measurement. To show this, we take a measurement at a detuning $\nu_0 \approx -1.25$ chosen arbitrarily, where there is no intrinsic TIN cancellation and the linearized measurement is strongly contaminated with TIN (blue trace in Fig.~\ref{fig:non_linear_readout}). In contrast to this, when the nonlinear transform of Eq.~\eqref{eq:non_linear_readout} is applied to the digitized photodetector output all orders of TIN are removed, improving the SNR in the recording of the high-Q mechanical mode by almost one order of magnitude (10 dB). The remaining imprecision noise is caused by the aforementioned broadband technical amplitude noise apparent in our system.\par

The procedure to obtain a calibrated mechanical spectrum is as follows. First, the photocurrent is converted into relative cavity occupation according to $n_\cav(t)/n_{\cav,0} = \qty[I(t)-I_\text{bg}]/(I_\text{max}-I_\text{bg})$, where $I_\text{max}$ is the photocurrent at $\nu=0$ and $I_\text{bg}$ is the background level. $I_\text{max}$ and $I_\text{bg}$ are estimated from a fast optical resonance scan such as the one presented in Fig.~\ref{fig:cavity_char}(c). Eq.~\eqref{eq:non_linear_readout} is then applied to obtain an estimate of the relative detuning fluctuations $\delta \nu(t)$. Finally, the apparent mechanical displacement of the high-Q mode $y$ (in meters) is calculated using: $y(t) = -\kappa \xzp \nu(t)/2 g_0$. The values of $\kappa=2\pi\cdot\SI{36}{\mega\hertz}$ and $g_0=2\pi\cdot\SI{441}{\hertz}$ are taken from the measurements shown in Fig.~\ref{fig:cavity_char}(b), whereas the zero-point fluctuation magnitude is $\xzp = \sqrt{\hbar/2 m_\text{eff} \Omega_m} =\SI{1.93}{\femto\meter}$ with the effective mass $m_\text{eff}=\SI{2}{\nano\gram}$ being estimated from a finite-element method calculation.

%% file: 6_conclusions.tex
We have investigated a membrane-in-the-middle microcavity that operates deeply in the nonlinear regime of transducing mechanical motion in optical amplitude or phase fluctuations. Utilizing an ultracoherent mechanical membrane, at room temperature our system is capable of  surpassing the threshold $C>n_\th$, where quantum backaction is stronger than thermal forces. However, the intracavity radiation pressure is dominated by an extraneous noise for cooperativities $C\gtrsim0.5n_\th$, preventing the onset of the quantum backaction dominated regime. We suspect this extra noise to be caused by cavity intrinsic frequency noise and if mitigated, third-order TINBA would be the dominating classical force. We furthermore characterize the origin of third order TIN components and introduce a post-processing step to realize TIN-free position readout.\par
Through removal of all orders of TIN from the measurement record, we provide a path for more efficient conditional cooling of mechanical motion~\cite{wieczorek_optimal_2015,magrini_real-time_2021}. Unconditional state preparation can also benefit from our protocol as its simplicity makes it suitable for real-time feedback systems. The nonlinear transform can be digitally implemented in a field-programmable gate array (FPGA) to increase the signal-to-noise ratio prior to an optimal control implementation~\cite{delic_cooling_2020-1}. As FPGAs allow for arbitrary transforms, a general nonlinear response could be corrected, such as the Fano-shaped response in sub-wavelength cavity optomechanics~\cite{fitzgerald_cavity_2021}. Beyond cavity optomechanics, the protocol can be directly applied to other systems possessing a Lorentzian response function. In particular, mechanically coupled Nitrogen-vacancy centers~\cite{arcizet_single_2011} and quantum dots~\cite{tanos_high-order_2024} have been shown to reach the nonlinear transduction regime. 

%% file: 7_acknowledgements.tex
We acknowledge support from the Danish National Research Foundation (bigQ, DNRF0142). We thank Yincheng Shi for helpful discussions and Angelo Manetta for help in setting up the experiment. We also thank the personnel at Vitrion for the deep etching of glass substrates and FiveNine Optics for the mirror coatings.